\documentclass[epj]{svjour}
\usepackage{graphics}
\begin{document}
\title{Communities recognition in the Chesapeake Bay ecosystem by dynamical clustering algorithms based on different oscillators systems}
\author{Alessandro Pluchino\inst{1}, Andrea  Rapisarda\inst{1} and Vito Latora\inst{1}
}                     
%
%
\institute{Dipartimento di Fisica e Astronomia,  Universit\'a di Catania, \\
and INFN sezione di Catania, Via S. Sofia 64,  I-95123 Catania, Italy}
\date{Received: date / Revised version: date}
%
\abstract{
We have recently introduced \cite{DC1,DC2} an efficient method for the detection and identification of modules in complex networks, based on the de-synchronization properties (dynamical clustering) of phase oscillators.
In this paper we apply the dynamical clustering tecnique to the identification of communities of marine organisms living in the Chesapeake Bay food web. We show that our algorithm is able to perform a very reliable classification of the real communities existing in this ecosystem by using different kinds of dynamical oscillators. We compare also our results with those of other methods for the detection of community structures in complex networks.
\PACS{
      {PACS-key}{discribing text of that key}   \and
      {PACS-key}{discribing text of that key}
     } 
} 
\maketitle
\section{Introduction}
\label{intro}

\begin{figure}
\begin{center}
\resizebox{0.75\columnwidth}{!}{%
  \includegraphics{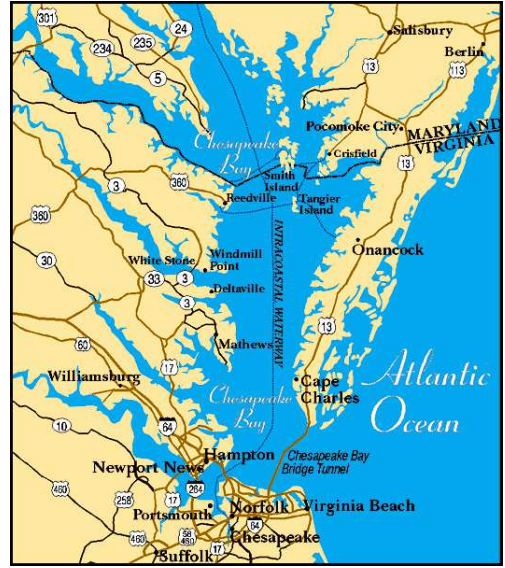}
}
\end{center}
\caption{Geographic position of the Chesapeake Bay ecosystem, situated on the Atlantic
coast of the United States.}
\label{chesapeake1}
\end{figure}

Complexity theory and associated methodologies are transforming ecological research, 
providing new perspectives on old questions as well as raising many new ones.
Patterns and processes resulting from interactions between individuals, populations, 
species and communities in landscapes are the core topic of ecology. 
These interactions form complex networks, which are the subject of intense research 
in complexity theory, informatics and statistical mechanics. This research has shown 
that complex natural networks often share common structures such as loops, trees and clusters, 
which contribute to widespread processes including feedback, non-linear dynamics, 
criticality and self-organization. 
\\
In ecosystems, and in particular in food webs, these structures have strong implications 
for their stability and dynamics. Actually, a food web constitutes a special description 
of a biological community with focus on trophic interactions between consumers and resources 
\cite{Ruiter}. Food webs are deeply interrelated with ecosystem processes 
and functioning since the trophic interactions represent the transfer rates of energy and matter 
within the ecosystem. In particular it is known that trophic webs are not randomly assembled, 
but are the result of the interaction of different cohesive subgroups ({\it modules} or 
{\it community structures}). Therefore, identifying the tightly connected groups within these networks 
is an important tool for understanding the main energy flows of the networks itself, as well as for 
defining a hierarchy of nodes and connections within a complex structure. 
\begin{figure}
\begin{center}
\resizebox{0.75\columnwidth}{!}{%
  \includegraphics{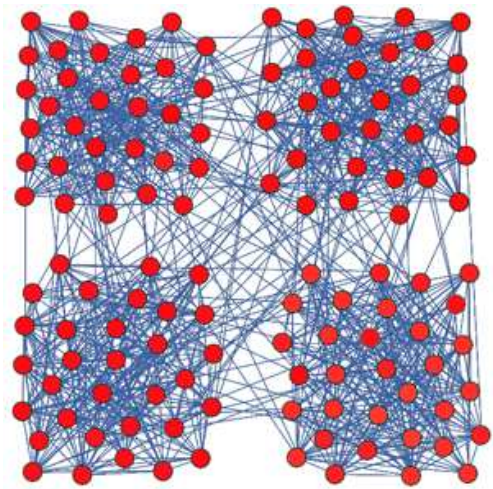}
}
\caption{An example of a network made of four communities or modules, defined as 
subsets of nodes {\it within} which the network connections (links) are dense,
but {\it between} which they are sparse. 
}
\label{network}
\end{center}
\end{figure}
\\
For practical purposes, modules can be defined as subsets of network nodes {\it within} 
which the connections are dense, but {\it between} which they are sparse (see Fig.\ref{network}).
In the last years many efficient heuristic methods have been proposed 
to investigate the presence of these structures in complex networks,
and their performances have been tested on both real and computer generated
networks with a known subdivision in different communities \cite{comparing,gudkov}. 
We have recently presented a {\it dynamical clustering} (DC) algorithm for 
modules identification based on the de-synchronization properties of a given 
dynamical system associated to the network \cite{DC1,DC2}. 
It combines topological and dynamical information
in order to recognize modular structures with a precision and 
a computational cost ($O(N^2)$ on a sparse graph) competitive with the majority 
of the other techniques.
In this paper we apply the DC algorithm to a well-known food web 
of marine organisms living in the Chesapeake Bay, situated on the Atlantic
coast of the United States (see Fig.\ref{chesapeake1}). 
We implement our algorithm  by using several dynamical systems
and we compare the results of the simulations among them 
and also with those obtained with other methods.

\section{Dynamics of weighted networks of coupled oscillators }
\label{sec:1}

The DC algorithm is based upon the well-known phenomenon of synchronization
of coupled phase oscillators \cite{reports}, each one
associated to a node of a given network, and interacting through the edges or links of the
graph.
In Ref.\cite{SPRL} it has been shown that an enhancement in the capability of synchronization can be achieved 
by using the information contained in the overall topology of the network.
This can be realized in practice through a weighting procedure wich associates 
a \textit{load} to each link of the network.
The load $l_{ij}$ of the link connecting nodes $i$ and $j$ can be quantified by
the so called {\it edge betweenness}, i.e. the fraction of shortest paths
that are making use of that link.
Within this assumption, the dynamics of a network of $N$ coupled oscillators $\{ {\bf x}_i \}_{i=1,...,N}$ is described by the following set of first order differential equations:
\begin{eqnarray}
\dot{\bf x}_i&=&{\bf F}({\bf x}_i)-\frac{\sigma}{\sum_{j\in K_i} ~ l_{ij}^\alpha} 
\sum_{j\in K_i} l_{ij}^\alpha ~{\bf H}[{\bf x}_i - {\bf x}_j]~~,\;\;\; 
\label{eq1}
\end{eqnarray}
where ${\bf F}={\bf F}({\bf x})$ governs the dynamics of each individual oscillator, 
${\bf H}={\bf H}({\bf x})$ is the coupling function, $\sigma$ is the overall coupling strength and
$K_i$ is the set of neighbors of node $i^{th}$.
Notice that the loads $\{l_{ij}\}$ have been raised to a power $\alpha$, where $\alpha$ is a real tunable parameter
which regulates the dynamical clustering process.
\\
In Ref.\cite{DC1} we showed that, for a given dynamical system ${\bf F}({\bf x}_i)$
and for a given value of the coupling strength $\sigma$, 
if the system starts in a perfectly synchronized state at $\alpha=0$ and $\alpha$ is let 
to slowly decrease in time from  $0$ to $-\infty$, the links with the higher load will be weighted
less and less with respect to the other links, thus inducing a progressive
desynchronization (dynamical clustering) of the system in a hierarchy of clusters of oscillators 
corresponding to different configurations of modules for the network considered.
In order to select which one of these configurations is the best one as
a function of $\alpha(t)$, we decided to look to local or global maxima 
of the \textit{modularity} $Q$. 
The latter simply compares the fraction of edges within $n_c$ arbitrary communities (intra-community links) 
of a given network with the expected fraction of such edges in a random network, which does not 
exhibits community structures \cite{NG}.
Actually, it is possible to define a $n_c \times n_c$ size matrix ${\mathbf
e}$ where the elements $e_{ij}$ represent the fraction of total links
starting at a node in partition $i$ and ending at a node in partition
$j$.  Then, the sum of any row (or column) of ${\mathbf e}$, $a_i
= \sum_j e_{ij}$ corresponds to the fraction of links connected to
$i$. If the communities were allocated without any regard to the underlying
structure, the expected number of intra-community links would be just $a_i \times a_i$. 
On the other hand, we know that the fraction of links exclusively within a partition is
$e_{ii}$. So, we can compare the two directly and sum over all the
partitions in the graph:
\begin{equation}
Q\equiv\sum_i(e_{ii} - a_i^2)
\label{defsq}
\end{equation}
Obviously it makes sense to look for high values of $Q$. In fact, if we take
the whole network as a single community, we get $Q=0$, while
values approaching $Q=1$ indicate strong community structure; 
on the other hand, for a random network we get again $Q=0$.
It is important to notice that the expression (\ref{defsq}) is not normalized, so that $Q$
cannot reach in practice the value $1$. 
For networks with an appreciable subdivision in classes, $Q$  
usually falls in the range from about $0.2$ to ~$0.7$. 
\\ 
In Refs.\cite{DC1,DC2} we applied the DC algorithm to several real and trial networks, 
using as dynamical systems the Opinion Changing Rate and the R\"ossler ones and adopting 
modularity $Q$ to choose the best subdivision for a given network. 
In the next sections, by using those and other dynamical systems - like the Kuramoto's one -, 
we will adopt again the modularity approach in order to explore the complex modular 
structure of the trophic relationships among organisms living in the Chesapeake Bay.

\begin{figure}
\begin{center}
\resizebox{1\columnwidth}{!}{%
  \includegraphics{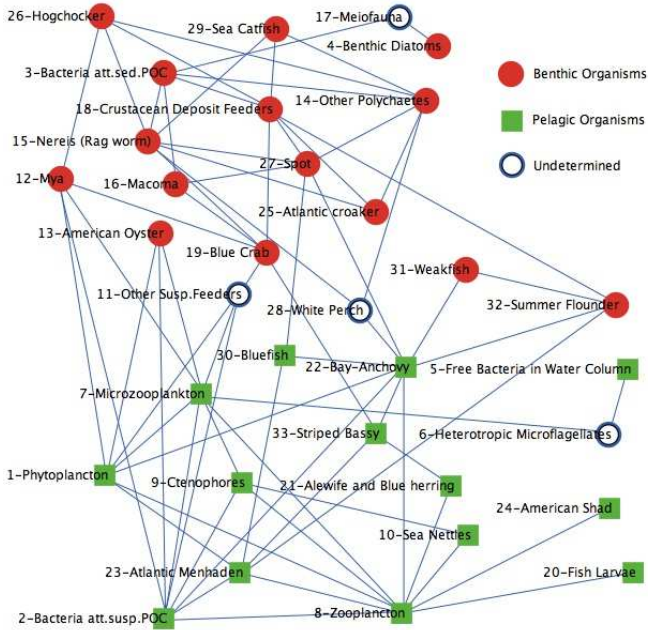}  
}
\end{center}
\caption{Chesapeake Bay food web. The nodes represent the 33 most important taxa, interconnected by 71 links (predatory interactions) and grouped in two main communities: the Benthic organisms (full circles) 
and the Pelagic organisms (full squares). Only four taxa (open circles) are undetermined. The modularity 
of such a natural subdivision is $Q=0.337$, if one considers nodes $17$ and $28$ as belonging to the 
Benthic community and nodes $6$ and $11$ to the Pelagic one. }
\label{chesapeake2}
\end{figure}

\section{Dynamical clustering analysis of the Chesapeake Bay food web}

The Chesapeake Bay watershed is a large and complex ecosystem of the U.S. Atlantic Coast, 
made up of smaller subsystems including forests, streams and marshes. 
Ecosystems work through the plants and animals that 
live in them. In a healthy ecosystem, plants and animals can benefit each other in a cycle 
of energy. Plants use solar energy to grow, transforming nutrients from the decay and waste 
of other living things. Animals eat the plants and recycle the nutrients, through their wastes 
and by their death and decay, for the use of other living things. The same process occurs on the 
land, in terrestrial ecosystems, and in the water, in aquatic ecosystems. Ecosystems continue 
to thrive when the energy from the nutrients in this cycle is not wasted or lost, but is stored 
and recycled.
\\
The Chesapeake Bay (CB) ecosystem was originally studied by Baird and Ulanowicz \cite{baird}, 
who carefully investigated the trophic relationships, i.e. the predatory interactions, 
between the 33 most important taxa (i.e. species or groups of species),
which can be represented by the nodes of a food network with 71 links (see Fig.\ref{chesapeake2}). 
Baird and Ulanowicz compiled the matrix of these relationships specifying the percentage of
carbon assimilated in each interaction, thus the network should be directed ($A$ feeds on $B$ but the
opposite is not true), and valued (due to the different
percentages of carbon exchanged in the interactions).
However, as usually done in many papers \cite{newgirv1,Eff},
we will consider it non-directed and non-valued.
\\
In Fig.\ref{chesapeake2}  the two main communities of organisms 
classified by Baird and Ulanowicz are visible: the Benthic ones (full circles), which live near the bottom of
the bay, and the Pelagic ones (full squares), which live near the surface or at middle depths).
Only four species (open circles) are undetermined.
We will verify in the following if the topological information about the predatory interactions, 
compressed in the edge betweennesses of the links of the CB food web (and stored in the load matrix $l_{ij}$, 
calculated once forever for this network), is sufficient to identify, by means of the dynamical clustering 
algorithm, these two main $a-priori$ communities, or similar configurations preserving in some way the
distinction between Benthic and Pelagic organisms. 
\\
Before going on, it is important to immediately calculate, by using the definition given in the 
previous section, the modularity $Q$ of the \textit{natural} subdivision (Benthic vs Pelagic) shown
in Fig.\ref{chesapeake2}. Actually, there are several possible configurations depending
on the arrangement of the nodes correspnding to undetermined species.
If we consider (as it seems more natural looking at the network itself) nodes $17$ and $28$ as belonging to the 
Benthic community and nodes $6$ and $11$ to the Pelagic community, the resulting modularity
is $Q=0.337$, a high value which indicates a good subdivision.
On the other hand, attributing node $11$ to the Benthic community and node $28$ to the Pelagic one
would produce a lower modularity score $Q=0.283$. Considering nodes $11$ and $28$ 
as a third separated community would  give back  a modularity $Q=0.321$. 
Therefore, assuming the configuration with two communities and $Q_{ref}=0.337$ as our reference 
configuration, in the next sections we apply our  DC algorithm to the Chesapeake Bay food web, using different oscillators' systems, in order to compare the resulting best configurations 
among them and with the one chosen as reference.

\subsection{Dynamical clustering with a system of R\"ossler oscillators}

The dynamics of a system of $N$ identical (three-dimen\-sional) chaotic R\"ossler oscillators, 
defined over the nodes of a given network, is ruled by Eq.(\ref{eq1}), with
${\bf x}_i = (x_i, y_i, z_i)$, 
${\bf F}({\bf x}_i) = [-\omega y_i - z_i, \omega x_i + 0.165 y_i, 0.2 + z_i(x_i - 10)]$
and ${\bf H}({\bf x})= [x, 0, 0] $ (thus the coupling acts only on the $x$ variable). 
In other words we have the following equations of motion \cite{DC2}:
\begin{displaymath}
\dot{x}_i=-\omega y_i - z_i-\frac{\sigma}{\sum_{j\in K_i} ~ l_{ij}^\alpha} 
\sum_{j\in K_i} l_{ij}^\alpha ~(x_i - x_j)  
\end{displaymath}
\begin{equation}
\dot{y}_i=\omega x_i + 0.165 y_i ~~~\;\;\;\;\;
\end{equation}
\begin{displaymath}
\dot{z}_i=0.2 + z_i(x_i - 10) ~~~\;\;\;\;\;\; i=1,\ldots ,N.
\label{rossler}
\end{displaymath}
Here $\omega$ is a common natural frequency associated
at each oscillator that, without loss of generality, we put equal to $1.0$. 
As previously seen, the load matrix $l_{ij}$ (the matrix of the edge betweennesses) 
is calculated once forever for the chosen network (in this case the CB food web).
\\
In order to evaluate the degree of synchronization of the R\"ossler system (\ref{rossler}) one 
has to calculate the order parameter 
$\Psi = \langle \frac{1}{N} \vert \sum_{i=1}^N e^{j\Phi_i (t)} \vert  \rangle_t$, 
where $\Phi_i (t)= arctan[\frac{y_i(t)}{x_i(t)}]$ indicates the istantaneous phase of
the $i$-th oscillator and $\langle . . .\rangle_t$ stays for a time average.
If all the oscillators rotate independently, no clusters exist and we have $\Psi \sim \frac{1}{\sqrt{N}}$.
On the contrary, if their motions are synchronized in phase, only one cluster exists and we obtain $\Psi \sim 1$.
Once a network is fixed, the first task is to find the value of the coupling parameter $\sigma$ 
providing a fully synchronized starting state for the R\"ossler oscillators at $\alpha=0$ (i.e. at $t=0$). 
Then, one can let $\alpha$ to decrease in time and study the dynamical clustering process
acting on the istantaneous phases $\Phi_i (t)$'s of the oscillators. 
We call "cluster" a group of contiguous phases in the $\Phi$'s interval
(usually $[-3,3]$) separated by a distance of more than $0.02$ units. 
For each value of $\alpha$ a different configuration of clusters (corresponding to a given network subdivision) 
will appear and one has to calculate the corresponding modularity and select
the configuration with the best modularity score.
\\
\begin{figure}
\begin{center}
\resizebox{0.9\columnwidth}{!}{%
  \includegraphics{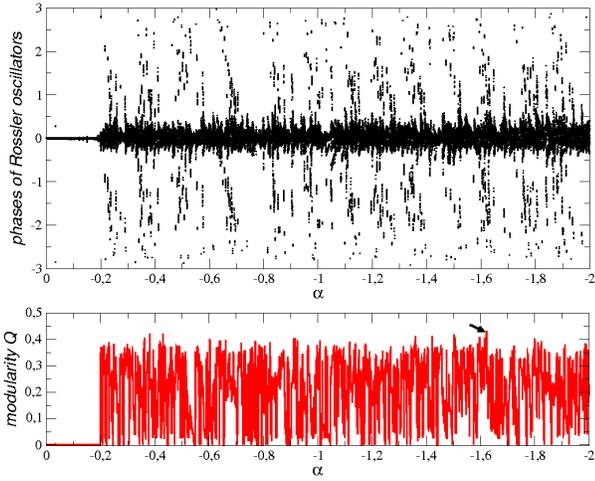}
}
\end{center}
\caption{A typical run of the DC algorithm for the Chesapeake Bay food web: time-evolution of both the R\"ossler's phases (top panel) and the corresponding modularity (bottom panel) as a function of $\alpha(t)$.
}
\label{rossler-dc}
\end{figure}
%
\begin{table}
\caption{For the R\"ossler case, clusters configuration with the best modularity score $Q_{best}=0.43$ at $\alpha_{best}=-1.62$, indicated with an arrow in the bottom panel of Fig.\ref{rossler-dc}.}
\label{tab-rossler} 
\begin{tabular}{ll}
\hline\noalign{\smallskip}
cluster & nodes   \\
\noalign{\smallskip}\hline\noalign{\smallskip}
cluster 1 (10 nodes) & 3,14,15,16,18,25,26,27,28,29\\
cluster 2 (3 nodes) & 4,17,19\\
cluster 3 (1 nodes) & 30\\
cluster 4 (3 nodes) & 22,31,32\\
cluster 5 (14 nodes) & 1,2,7,8,9,10,11,12,13,20,21,23,24,33\\
cluster 6 (2 nodes) & 5,6\\
\noalign{\smallskip}\hline
\end{tabular}
\end{table}
In Fig.\ref{rossler-dc} we show the result of a typical event of the R\"ossler DC algorithm
for a value of the interaction strenght $\sigma=1$ (such that the system would lie
in its synchronized phase for $\alpha=0$).
The clusters evolution (top panel) and the corresponding modularity $Q(t)$ (bottom panel)
are plotted as a function of $\alpha$ 
(note that, in the top panel, the average istantaneous phase of
the system has been subtracted from the istantaneous phases of the oscillators 
in order to have a symmetric plot).
The system starts in a fully synchronized state ($x_i(0)=y_i(0)=z_i(0)=0$ $\forall i$)
at $\alpha_{start}=0$ and evolves through decreasing values of $\alpha(t)$ 
(with a decrement $\delta\alpha=0.0008$), up to the value
$\alpha_{end}=-2$. Even if the system strongly oscillates during the 
desynchronization process, clusters' configurations (i.e. community structures of the 
underlying network) with large values of modularity appear.
The detailed configuration with the highest modularity peak (see the arrow in the bottom panel) 
is reported in Table \ref{tab-rossler}. It consists of 6 clusters with a $Q_{best}=0.43$, obtained  for $\alpha_{best}=-1.62$, and it is quite consistent with the distinction between pelagic organisms 
(clusters n.1,2 and 4) and benthic organisms (clusters n.3,5 and 6).
Furthermore, if compared with the reference configuration, where $Q_{ref}=0.337$, the configuration we found
here seems evidently (having a higher modularity) to better reflect the underlying structure emerging from the global information stored in the edge betweennesses of the food web.

\subsection{Dynamical clustering with the Kuramoto model}

The Kuramoto model describes a population of $N$ periodic
oscillators having natural frequencies $\omega_i$ and
coupled through the sine of their phase differences \cite{kuramoto_model}.
It is simple enough to be analytically solvable, still retaining 
the basic principles to produce a rich variety of dynamical regimes
and synchronization patterns \cite{k1,strogatz}. 
\begin{figure}
\begin{center}
\resizebox{0.9\columnwidth}{!}{%
  \includegraphics{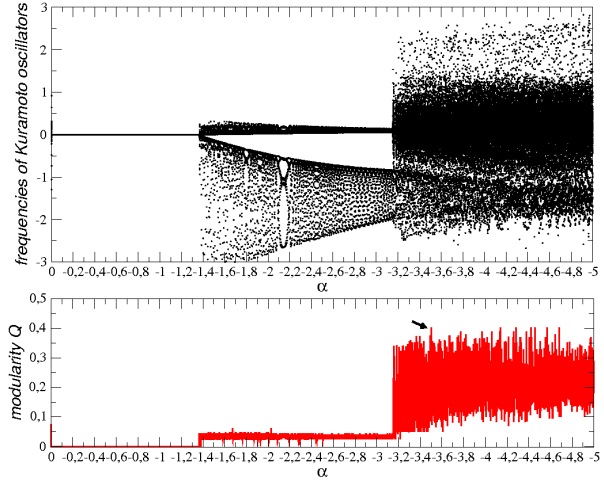}
}
\end{center}
\caption{A typical run of the DC algorithm for the Chesapeake Bay food web: time-evolution of both the Kuramoto's frequencies (top panel) and the corresponding modularity (bottom panel) as a function of $\alpha(t)$.
}
\label{kuramoto-dc}
\end{figure}
\\ 
The dynamics of the model is given by
\begin{equation}
    \dot{\theta_i} (t)  = \omega_i + \frac{K}{N} \sum_{j=1}^N
      \sin ( \theta_j  - \theta_i )  ~~~~~i=1,\dots,N
\label{kuramoto_eq1}
\end{equation}
where $\theta_i (t)$ is the phase (angle) of the $i$th oscillator at time $t$,
while $\omega_i$ is its intrinsic frequency randomly drawn from
a symmetric, unimodal distribution $g(\omega)$ with a first moment
$\omega_0$ (typically a Gaussian distribution or a uniform one). These
natural frequencies $\omega_i$ are constant and time-independent. 
The sum in the above equation is running over all the
oscillators so that this is an example of a globally coupled system. 
The most interesting feature of the model is that,
despite the difference in the natural frequencies of the
oscillators, it exhibits a
spontaneous transition from incoherence to collective synchronization
beyond  a certain threshold $K_c$ of the coupling strength $K$ \cite{strogatz}.
For small (positive) values of $K$, each oscillator tends to run independently with its own
frequency,  while for large values of  $K$, the coupling tends to synchronize (in phase and frequency)
the oscillator with all the others.
\\
In the Kuramoto model,  it  is possible to define a complex mean field
order parameter as
\begin{equation}
r e^{i \Psi} = \frac{1}{N} \sum_{j=1}^N   e^{i \theta_j}~~~,
\label{kuramoto_orderparameter}
\end{equation}
where the magnitude $0 \le r(t) \le 1$ is a measure of the
coherence of the population and $\Psi(t)$ is the average
phase. In other words, as $N$ approaches infinity, the magnitude $r_\infty$the average istantaneous phase of
the system has been subtracted from the istantaneous phases of the oscillators 
in order to have a symmetric plot)
of the complex mean field after a transient time should be zero in
the incoherent state with $K \le K_c$ and different from zero
in the coherent state with $K > K_c$. 
Actually, as $K$ increases  beyond $K_c$, more and more 
oscillators will be recruited toward the mean phase $\Psi(t)$
and $r_\infty$ is expected to continuously increase from zero
to one.

\begin{table}
\caption{Fot the Kuramoto case, clusters configuration with the best modularity score $Q_{best}=0.404$ at $\alpha_{best}=-3.50$,
indicated with an arrow in the bottom panel of Fig.\ref{kuramoto-dc}.}
\label{tab-kuramoto}
\begin{tabular}{ll}
\hline\noalign{\smallskip}
cluster & nodes  \\
\noalign{\smallskip}\hline\noalign{\smallskip}
cluster 1 (2 nodes) & 4,17\\   
cluster 2 (2 nodes) & 9,10\\
cluster 3 (11 nodes) & 3,14,15,16,18,19,25,26,27,28,29\\
cluster 4 (14 nodes) & 1,2,7,8,11,12,13,20,21,22,23,24,30,33\\
cluster 5 (2 nodes) & 5,6\\
cluster 6 (2 nodes) & 31,32\\ 
\noalign{\smallskip}\hline
\end{tabular}
\end{table}

In order to use Kuramoto model as dynamical system in the context
of our dynamical clustering algorithm,
we immediately note that, at variance with the R\"ossler system,  
Eqs.\ref{kuramoto_eq1} are already very similar
to Eqs.\ref{eq1}, providing that one considers  
${\bf x}_i = \theta_i$, ${\bf F}({\bf x}_i) = \omega_i$,
$\sigma=K$, $l_{ij}=1$ and ${\bf H}({\bf x})= \sin ( \theta_j  - \theta_i )$.
In other words Kuramoto equations already have a coupling term, containing
a nonlinear function that becomes approximatively linear in the synchronization manifold 
(where $\theta_j \sim \theta_i)$).
This makes Kuramoto model particularly suitable for our purpose.
In fact in this case, after having chosen a given network and having associated
an oscillator to each node, we can directly apply the 
weighting procedure of section 2 (see Eq.\ref{eq1}) to Eqs.\ref{kuramoto_eq1} without
adding further coupling terms, thus simply obtaining:
\begin{equation}
    \dot{\theta_i} (t)  = \omega_i + \frac{K}{\sum_{j\in K_i} ~ l_{ij}^\alpha} 
\sum_{j\in K_i} l_{ij}^\alpha  \sin ( \theta_j  - \theta_i )  ~~~~~i=1,\dots,N
\label{kuramoto_weighted}
\end{equation}
where as usual, $l_{ij}$ is the load of the link connecting nodes $i$ and $j$ in the chosen network,
$K_i$ the set of neighbors of node $i^{th}$ and $\alpha$ is  a real tunable parameter. 
\\
As in the previous section, our task is now to test the sensitivity of the Kuramoto
dynamical clustering algorithm on the CB food network. 
First of all, we have to fix the coupling
parameter $K$ of Eq.\ref{kuramoto_weighted} in order to obtain a fully synchronized state
of the system for $\alpha\sim0$. We found that for $K>5$ such a state is guaranteed,
thus we will reasonably set $K=10$.          
In our simulations of the Kuramoto system we will always use as initial conditions 
uniform distributions for both the $\theta_i$'s (in the interval $[-\pi,\pi]$) and 
$\omega_i$'s (in the interval $[-2,2]$). We remind that the latter are constant in time. 
At variance with the R\"ossler case, we are now interested to the istantaneous
frequencies $\dot{\theta_i} (t)$ of the oscillators (which at $t=0$ coincide with the
natural frequencies $\omega_i$). Again, the average istantaneous frequency of
the system will be subtracted from the istantaneous frequencies of the oscillators 
in order to have a symmetric plot.
\\
In Fig.\ref{kuramoto-dc} we show a typical run of the Kuramoto DC algorithm for the CB food network.
As before, the simulation starts from $\alpha_{start}=0$ then 
$\alpha$ decreases in time with a given step ($\delta\alpha=0.0008$): one sees that in a few steps the system
suddenly synchronizes (due to the high value of $K$) then slowly relaxes producing a 
progressive desynchronization that is, again, very oscillating in time; for each value of $\alpha(t)$ 
the istantaneous clusters' configuration of frequencies is identified
(being the definition of 'cluster' the same than in the previuos section), 
and the correspondent modularity calculated.
The detailed configuration with the highest modularity peak (see the arrow
in the bottom panel) is reported in Table \ref{tab-kuramoto}.
It consists of 6 clusters with a $Q_{best}=0.404$, obtained  for $\alpha_{best}=-3.50$:
even if this value of modularity is less than in the R\"ossler case, on the other hand it is
greater  than $Q_{ref}=0.337$ and in any case is again quite consistent with the distinction 
between pelagic organisms and the benthic ones.

\subsection{Dynamical clustering with the Opinion Changing Rate model}

As a last example of application of the dynamical clustering algorithm to the Chesapeake Bay food web, 
let us to consider as dynamical system the so called Opinion Changing Rate (OCR) model \cite{DC1}. 
It was originally introduced in Ref.\cite{ocr} as a modification of the Kuramoto model,
in order to study how the personal inclination to change, randomly  distributed in a group of individuals,  can affect the opinion dynamics of the group itself.
The dynamics of a system of $N$ fully coupled individuals (oscillators) is governed by 
the following set of differential equations:
\begin{equation}
    \dot{x_i} (t)  = \omega_i + \frac{\sigma}{N}\sum_{j}
      ~\beta \sin ( x_j  - x_i ) e^{- \beta |x_j  - x_i| }  ~~~~~i=1,\dots,N
\label{OCR_eq1}
\end{equation}
where $x_i (t)$ is the opinion of the $i$th agent at time $t$. 
Here the opinions have a very general meaning and can be usefully represented 
by means of unlimited real numbers 
$x_i \in ] -\infty +\infty[ ~~ \forall i=1,...,N$.
Opinions interact by means of the coupling term, where
$\sigma$ is the coupling strength and
the exponential factor, tuned by the parameter $\beta$,
ensures that opinions will not influence each other any longer 
when the reciprocal distance exceeds a certain threshold.
This is perhaps the most remarkable feature of the OCR model, 
since it allows the system to reach an asymptotic stationary state 
where the configuration of opinions does not vary any longer.
The parameter $\beta$ appears also as a factor of the sine in the coupling term and
simply rescales the range of the coupling strenght.
We typically adopted the value $\beta$=3, which ensures a consistent behavior
of the exponential decay.
Finally, the $\omega_i$'s - corresponding to the {\it natural frequencies} of the 
oscillators in the Kuramoto model - represent here the so-called
\textit{natural opinion changing rates}, i.e.
the intrinsic inclinations of the agents
to change their opinions. 
For this reason we called this model the {\it Opinion Changing Rate}
(OCR) model \cite{ocr}. 
The values $\omega_i$'s, which do not depend on time, are uniformly distributed 
in the range $[-0.5,0.5]$ and represent also the initial conditions for the 
istantaneous frequencies $\dot{x_i}(t)$'s. 
\\
Many numerical simulations were performed in Ref.\cite{ocr} 
starting from a uniform distributions of the
initial opinions $x_i(t=0)$ in the range $[-1,1]$.
As a function of the coupling strength $\sigma$, a transition was observed 
from an incoherent phase (for $\sigma < \sigma_c$), in which people 
tend to preserve different opinions and different frequencies according to their natural 
changing rates $\omega_i$, 
to a partially synchronized phase, where people share a small number of opinions, 
and, finally, to a fully synchronized one (for $\sigma >> \sigma_c$) in which all the people 
share the same opinion changing with the same rate.
In order to measure the degree of synchronization of the system,
it can be adopted an order parameter related to the standard deviation of the
istantaneous frequencies and defined as 
$R(t) = 1 - \sqrt{ \frac{1}{N} \sum_{j=1}^N (\dot{x}_j(t) - \dot{X}(t))^{2}}$,
where $\dot{X}(t)$ is the average over all individuals of $\dot{x_j}(t)$.
From such a definition it follows that
$R=1$ in the fully synchronized phase and $R<1$ in the incoherent or 
partially synchronized phase. 
\\
\begin{figure}
\begin{center}
\resizebox{0.9\columnwidth}{!}{%
  \includegraphics{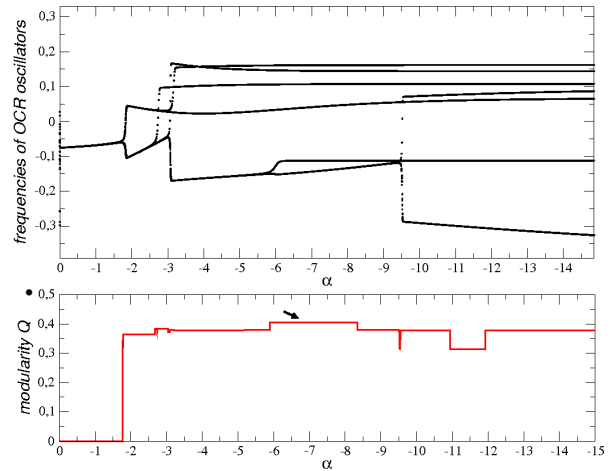}  
}
\end{center}
\caption{A typical run of the DC algorithm for the Chesapeake Bay food web: time-evolution of both the OCR's frequencies (top panel) and the corresponding modularity (bottom panel) as a function of $\alpha(t)$.
}
\label{ocr-dc}
\end{figure}
\begin{table}
\caption{For  the OCR  model, clusters configuration with the best modularity score $Q_{best}=0.404$ at $\alpha_{best}\sim-5.89$, indicated with an arrow in the bottom  panel of Fig.\ref{ocr-dc}.}
\label{tab-ocr}
\begin{tabular}{ll}
\hline\noalign{\smallskip}
cluster & nodes   \\
\noalign{\smallskip}\hline\noalign{\smallskip}
cluster 1 (14 nodes) & 1,2,7,8,11,12,13,20,21,22,23,24,30,33\\   
cluster 2 (2 nodes) & 31,32\\    
cluster 3 (11 nodes) & 3,14,15,16,18,19,25,26,27,28,29\\
cluster 4 (2 nodes) & 5,6\\   
cluster 5 (2 nodes) & 9,10\\
cluster 6 (2 nodes) & 4,17\\
\noalign{\smallskip}\hline
\end{tabular}
\end{table}

\subsubsection{Standard Opinion Changing Rate model}

In order to utilize the OCR model as dynamical system for recovering community structures
in the CB food network we put togheter Eq.\ref{eq1} and Eq.\ref{OCR_eq1}, thus obtaining  
\begin{equation}
    \dot{x_i} (t)  = \omega_i + \frac{\sigma}{\sum_{j\in K_i}~ l_{ij}^\alpha} 
    \sum_{j\in K_i}  ~\beta ~ l_{ij}^\alpha \sin ( x_j  - x_i ) e^{- \beta |x_j  - x_i| } , 
\label{OCR_eq2}
\end{equation}
where $i=1,\dots,N$, $\alpha$ is the usual real tunable parameter and $K_i$ is the set of 
neighbors of node $i^{th}$. As in the case of Kuramoto model, we
do not need any further coupling term in Eqs. \ref{OCR_eq2}, since
such a term is already present in the OCR model.
We will follow now the time evolution of the istantaneous frequencies $\dot{x_i} (t)$ 
starting, as usual, from a completely synchronized state at $\alpha\sim0$:
more precisely, as in the Kuramoto case, the initial frequency distribution is uniform inside 
the interval $[-0.5,0.5]$ (since it coincides with the $\omega_i$'s distribution) 
but we chose $\sigma=5.0$, a coupling that ensures a rapid synchronization for $t>0$.
Then we let $\alpha$ to decrease in time with a step $\delta\alpha=0.01$. 
We are confident that, at variance with the previous examples, the exponential factor 
in the coupling term could trigger the aggregation of frequencies in stable clusters 
corresponding to community configurations of the network. 
\\
In Fig.\ref{ocr-dc} we plot the OCR frequencies' time evolution for the
Chesapeake Bay food web, together with the respective modularity.
One can immediately appreciate the stability of the desynchronization process,
which produces (due to the exponential factor in the coupling term) a branching sequence 
of metastable plateaux corresponding to different clusters configurations. 
The best one is obtained for $-5.9 > \alpha_{best} > -8.3$ with a modularity 
$Q_{best}=0.404$ and corresponds exactly to the same best configuration (six clusters) 
of the Kuramoto case, see Table\ref{tab-ocr}. 
But the stability of the synchronized manifold makes useless, in this case, 
to adopt lower values of $\delta\alpha$, thus making also the simulation
less expensive in terms of computational cost.

\subsubsection{Opinion Changing Rate model with HK dynamics}

\begin{figure}
\begin{center}
\resizebox{0.9\columnwidth}{!}{%
  \includegraphics{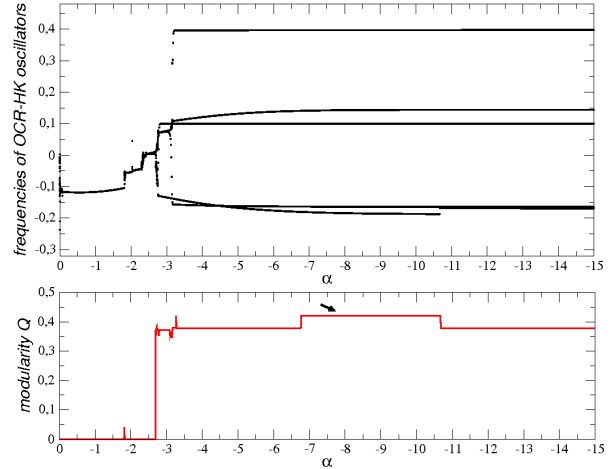}
}
\end{center}
\caption{A typical run of the DC algorithm for the Chesapeake Bay food web: time-evolution of both the OCR-HK's frequencies (top panel) and the corresponding modularity (bottom panel) as a function of $\alpha(t)$.
}
\label{ocr-hk-dc}
\end{figure}

In order to further improve the performances of the OCR system, in Ref.\cite{DC1} we tought to modify 
the natural opinion changing rates $\omega$'s following a suggestion from the
Hegselmann and Krause (HK) model of opinion formation. 
The HK model \cite{HK} is based on the concept of bounded
confidence, i.e. on the presence of a parameter $\epsilon$, called
{\it confidence bound}, which expresses the compatibility among
the agents in the {\it opinion space}. If the opinions of two
agents $i$ and $j$ differ by less than $\epsilon$, their positions
are close enough to allow for a discussion, which eventually leads
to a change in their opinions, otherwise the two agents do not
interact with each other. In our case, the opinion space is one-dimensional,
being usually $x_i(t) \in [-1,1]$. Thus we fixed a small value for $\epsilon$ 
and we let the $\omega$'s to change in time (while in the standard OCR version the $\omega$'s
are time independent) starting from a random uniform distribution in the interval $[-0.5,0.5]$ and
according to the following HK dynamics: at each time step
a given agent, with an opinion $x_i$ and a natural frequency $\omega_i$, checks how many of 
its neighbors (according to the network topology) 
are compatible, i.e. lie inside the confidence range [$x_i-\epsilon,x_i+\epsilon$] in the
opinion space. Next, the $\omega_i(t)$ of the agent takes the average value of the 
$\omega$'s of its compatible neighbors at time $t-1$.
We will refer to this new system as $OCR-HK$. 
In the OCR-HK system the secondary process acting on the $\omega$'s is superimposed to the main
dynamical evolution of Eq.\ref{OCR_eq2} and contributes to further stabilize the  
istantaneous frequencies $\dot{x_i}(t)$ of the agents (oscillators).
\\
In Fig.\ref{ocr-hk-dc} the new simulation with the OCR-HK system are shown for the Chesapeake Bay food web. 
This simulation was performed again with $\sigma=5.0$ 
and a $\delta\alpha=0.01$, with a confidence bound $\epsilon=0.0005$.
Again a branching desynchronization process with well defined metastable 
plateaux occurs.
But in this case, as expected, the $HK$ dynamics produces a highest value of 
modularity, $Q_{best}=0.42$ (very near to that obtained with the R\"ossler algorithm),
which is reached for $-6.77 > \alpha_{best}  > -10.67$, yielding
a subdivision of the food web into five communities, whose detailed structure
is shown in Table\ref{tab-ocr-hk}. 
It clearly appears that, besides the greater value of $Q$, the distinction between pelagic and benthic 
organisms is improved with respect to the standard OCR case too.
In any case, both the standard OCR and the OCR-HK algorithm give a best configuration
whose modularity is higher than the reference one $Q_{ref}=0.337$.

\begin{table}
\caption{For the OCR-HK case, clusters configuration with the best modularity score $Q_{best}=0.42$ at $\alpha_{best}\sim-6.77$,indicated with an arrow in the bottom panel of Fig.\ref{ocr-hk-dc}.}
\label{tab-ocr-hk}
\begin{tabular}{ll}
\hline\noalign{\smallskip}
cluster  & nodes   \\
\noalign{\smallskip}\hline\noalign{\smallskip}
cluster 1 (2 nodes) & 31,32\\      
cluster 2 (11 nodes) & 3,14,15,16,18,19,25,26,27,28,29\\
cluster 3 (2 nodes) & 4,17\\ 
cluster 4 (14 nodes) & 1,2,7,8,9,10,11,12,13,20,21,22,23,24,30,33\\
cluster 5 (2 nodes) & 5,6\\
\noalign{\smallskip}\hline
\end{tabular}
\end{table}

\section{Discussion and Conclusions}

Summarizing, the dynamical clustering (DC) algorithm, which exploits the synchronization
properties of some dynamical system of oscillators associated with the nodes of a given
complex network, seems to work very well in identifying the underlying community structure 
of the Chesapeake Bat food web of trophic relationships.
In fact we shown that, whatever the dynamical system we use (R\"ossler, Kuramoto, OCR
or OCR-HK), the DC algorithm discovers community configurations with high values of
modularity, in all the cases higher than the reference configuration of Fig.\ref{chesapeake2} 
related to the main subdivision of the Chesapeake Bay network in Benthic and Pelagic organisms.
This would imply also that, if we consider modularity optimization as a valid method to retrieve the best 
community structure compatible with the information stored in the topology of a given network, 
then we should conclude that the rigid subdivision in Benthic and Pelagic organisms is not the 
optimal one for the Chesapeake Bay food web. 
Actually, modularity method has been recently questioned by \cite{Fortunato-PNAS}, which
showed that it is $a-priori$ impossible to tell whether a module, detected through modularity
optimization, is indeed a single module or a cluster of smaller modules, but the problem
is still open.
In any case it is worthwhile to notice that the DC algorithm produces better results 
also if compared with other methods for detecting community structures using modularity 
optimization, which in the past were applied to the Chesapeake Bay (CB) food web.
For example in Ref.\cite{NG}, where Girvan and Newman presented their original iterative 
method, based in finding and removing progressively the edges with the largest
betweenness until the network breaks up into its components, the best modularity value
for the CB food web was $Q_{GN}=0.380$; and in Ref.\cite{Eff}, where another 
method based on the Information Centrality, was proposed and applyied to the CB food web, 
the best score for the modularity was $Q_{IC}=0.376$.
\\
In conclusion, the DC algorithm, in particular in the version using the OCR-HK system 
(which produces large modularity configurations with very stable synchronization
patterns), seems to be a very efficient method for the study of community structures 
in ecosystems and food webs, even if in the approximation of non-directed networks. 
In this direction, a further improvement in the DC algorithm 
performance probably could come from the use of a recent generalization of the modularity approach 
which incorporate also the information contained in edge directions \cite{Leicht-Newman}.

\end{document}